\documentstyle[12pt]{article}

\def\nfrac#1#2{{\displaystyle{\vphantom1\smash{\lower .5 ex\hbox{\small$#1$}}%
	\over\vphantom1\smash{\raise .25 ex\hbox{\small$#2$}}}}}

\begin{document}

\begin{titlepage}

\begin{center}
May 5, 1994\hfill    TAUP--2155--94 \\
\hfill    hep-th/9405039

\vskip 2 cm

{\large \bf
The $N=1$ superstring as a topological field theory.
}

\vskip 1 cm

Neil Marcus\footnote[1]{
Work supported in part by the US-Israel Binational Science Foundation,
and the Israel Academy of Science.
E-Mail: NEIL@HALO.TAU.AC.IL}

\vskip 0.2 cm

{\sl
School of Physics and Astronomy\\Raymond and Beverly Sackler Faculty
of Exact Sciences\\Tel-Aviv University\\Ramat Aviv, Tel-Aviv 69978, ISRAEL.
}

\end{center}

\vskip 4 cm

\begin{abstract}

\noindent

By ``untwisting'' the construction of Berkovits and Vafa~\cite{berk}, one can
see that the $N=1$ superstring contains a topological twisted $N=2$ algebra,
with central charge $\hat c = 2$.  We discuss to what extent the superstring is
actually a topological theory.

\end{abstract}

\end{titlepage}

\parskip 6 pt
\parindent 2 em

Berkovits and Vafa (BV) have shown that the $N=1$ Neveu-Schwarz-Ramond (NSR)
string can be viewed as an $N=2$ string~\cite{berk}.  They do this by twisting
the stress tensor of the theory with a $U(1)$ ghost current:
$T \rightarrow T -\frac12 \, \partial J$, so that the $(b,c)$ and
$(\beta,\gamma)$ ghosts of the string become
dimension $\left(\frac32,-\frac12\right)$ and $(1,0)$,
respectively\footnote{Our conventions are those of Friedan, Martinec and
Shenker~\cite{FMS}.  With respect to BV, we redefine $\gamma \rightarrow
-\gamma/2$, $\beta \rightarrow -2 \beta$, and $J \rightarrow -J$.  As usual, we
bosonize the $(\beta,\gamma)$ system by defining $\gamma = \eta \, {\rm
e}^\phi$,
and $\beta = \partial \xi \, e^{-\phi}$, noting that $\partial \phi = \beta
\gamma$.  We consider only the holomorphic sector of the theory, and shall
often leave the $z$ dependence implicit, where this is not confusing.}.  The
theory then contains an $N=2$ superconformal algebra generated by the twisted
stress tensor $T(z)$, the $U(1)$ ``R--parity''
current $J(z)$, the twisted ghost $b(z)$, and
an improved BRST current $Q(z)$.  The NSR matter has an $N=2$ central charge
$\hat c = 5$, and the twisted ghosts $\hat c = -3$, so the theory is now
a critical
$N=2$ matter theory with $\hat c = 2$.  The final step of their construction is
to make an $N=2$ string by coupling this matter to an $N=2$ ghost
system.  It can be shown that this new string is equivalent to the original NSR
string~\cite{berk,equiv}.

Having an $N=2$ algebra, it is natural to ask what happens if one further
twists it to obtain a ``topological $N=2$ algebra''~\cite{twisted}.
Such a twisting can be done in two inequivalent ways, $T \rightarrow T \pm
\frac12 \, \partial J$, since the theory is not in a unitary representation of
the untwisted $N=2$ algebra ($b \neq Q^\dagger$).  The negative twist gives a
doubly-twisted NSR string, which we shall not consider further.  The positive
twist undoes the twisting of BV, restoring the stress tensor to that of the
original theory.  The generators of the topological $N=2$ algebra are then:
\begin{eqnarray}
\label{ops}
B &=& b \ , \nonumber \\
J &=& c \, b + \eta \, \xi \ , \nonumber \\
T &\rightarrow& T + \nfrac12 \, \partial J \nonumber \\[-2 pt]
  &=& T_m - \nfrac32 \, \beta \, \partial\gamma
	- \nfrac12 \, \partial\beta \, \gamma
	- 2 b\, \partial c - \partial b\, c \\[2 pt]
  &=& T_{N=1} \ , \nonumber \\[2 pt]
Q &=& -\nfrac12 \, \gamma \, G_m +
	c \left( T_m - \nfrac32 \, \beta \, \partial\gamma
	- \nfrac12 \, \partial\beta \, \gamma \right) + b\,c\,\partial c
	-\nfrac14 \, \gamma^2 b
	+ \partial (c\,\xi \eta)
      + \partial^2 c \nonumber \\
  &=& Q_{N=1} + \partial \left( c\,\xi \eta + \partial c
                  -\nfrac32 \, c \, \partial \phi \right) \ , \nonumber
\end{eqnarray}
and the algebra is described by the nontrivial operator product expansions:
\begin{eqnarray}
\label{alg}
Q(z)\cdot B(w) & \sim & { {\hat c}\over (z-w)^3}+{J(w)\over{(z-w)^2}} +
	{T(w) \over (z-w)}\ , \nonumber \\
T(z)\cdot J(w) & \sim & - \, {{\hat c}\over{(z-w)^3}}
	+ {J(w)\over{(z-w)^2}}+{\partial J(w)\over (z-w)}\ , \\
J(z)\cdot J(w) & \sim & {\hat c \over{(z-w)^2}} \ ,\nonumber
\end{eqnarray}
with $\hat c = 2$, with the remaining OPE's expressing the fact that $B(z)$ and
$Q(z)$ are nilpotent; that $T$, $B$ and $Q$ are fields with conformal
spin 2, 2 and 1, respectively;  and that $Q$ and $B$ have R--parity
$\pm 1$, respectively.  These additional OPE's have no central terms.  The
algebra can easily be checked explicitly: the only difficult OPE, $Q(z) \cdot
Q(w) \sim 0$, is a consequence of $Q^2 = 0$ and $J(z)\cdot Q(w) \sim Q(w) /
(z-w)$.

Note that the stress tensor $T(z)$ and the BRST charge
$Q$---but not the BRST current $Q(z)$---of~(\ref{ops}) are those of the
original NSR string.  The operator algebra implies that $T(z)$ and $Q(z)$ are
both BRST exact:
\begin{eqnarray}
\label{exact}
T(z) & = & \left\{\, Q \, , \, B(z) \, \right\} \ , \nonumber \\
Q(z) & = & - \left[\, Q \, , \, J(z) \, \right] \ ,
\end{eqnarray}
which can be considered to be the  defining relations of a topological
conformal field theory~\cite{DVV}.  As is well known, even the bosonic string
is topological in some sense, since it satisfies~(\ref{exact}) with an
unimproved $Q(z)$ and $J$ replaced by the ghost-number current
$J_{gh}$~\cite{DVV}.  However, the algebra of these currents does not close, so
one does
not have the full twisted $N=2$ algebra\footnote{In $N \ge 2$
strings the algebra does close, giving rise directly
to a topological twisted $N+2$
algebra~\cite{N=2->4}.}.  To close the algebra one needs to find some $U(1)$
current to add to $J_{gh}$, so that the improved $Q(z)$, now {\it defined\/}
by~(\ref{exact}), has the correct OPE with $J(w)$.  The twisted $N=2$ algebra
then
follows.  This procedure has been carried out for the bosonic
string~\cite{alltop1,alltop2}, and for the NSR string~\cite{alltop2}, where a
larger
twisted $N=3$ algebra is found.  In these constructions the extra $U(1)$
current is taken from the matter sector of the theory---for example one can use
the momentum current of the Liouville field---while
in our case $J$ is defined purely
from the ghost fields.  This has the advantage that our construction is valid
for any NSR background (with central charge $c=10$), since it does not require
the existence of a matter $U(1)$ current.  Also, since we have not interfered
with the matter sector at all, we do not break Lorentz invariance in the flat
Minkowski background.

The surprising feature of our construction---and the
reason that it did not appear among the topological algebras of
ref.~\cite{alltop2}---is that the bosonization of the $(\beta,\gamma)$ ghost
system, which is basically a calculational tool in usual NSR calculations, is
crucial here in defining $J$ and $Q$.  (This is, of course, also true in the BV
embedding of the NSR string into the $N=2$ string, since we have borrowed $J$
from them.) To see this,  note that $J = J_{gh} - J_P$, where $J_{gh}$ is the
full ghost number current, and $J_{P}$ is the ``picture-number'' current:
\begin{eqnarray}
\label{picture}
J_P &=& - \beta \gamma - \eta \, \xi \nonumber\\
    &=& - \partial \phi - \eta \, \xi \ .
\end{eqnarray}
Since $J_P$ is the difference between the $(\beta,\gamma)$ ghost-number current
and the $(\xi,\eta)$ fermion-number current, its charge commutes with all the
fields of the pre-bosonized NSR string, and it can not be written in terms of
them.   Another indication of how fundamental is the bosonization, is that no
appropriate $J(z)$ can be defined in the manifestly supersymmetric
bosonization~\cite{super-bos}, or in the bosonization in which the roles of
$\eta$
and $\partial \xi$ are interchanged~\cite{other-bos}.

Since one now has a topological algebra in the NSR string, one can ask whether
the string
can be recast into a topological conformal field theory (TCFT).  To do this,
one needs to show that both the states and the amplitudes in the TCFT approach
agree with those of the NSR string.  Since states are given by the cohomology
of the BRST charge both in TCFT's and in regular strings, and since the BRST
charges of the two theories are identical, it appears to be obvious that one
obtains the correct states in a TCFT approach.  However, there are two somewhat
intertwined complications which need to be dealt with:  First, in TCFT's one
usually demands that states in the cohomology be annihilated by all the $B_n$'s
with $n > 0$~\cite{DVV}, implying that the Virasoro generators $L_n$ also
annihilate the states.  In twisted $N=2$ theories one also desires that states
in the cohomology be the twistings of chiral primary states in the original
$N=2$.  Such states must satisfy the additional requirement that they be
annihilated by $B_0$, and by the $J_n$'s and $Q_n$'s with $n > 0$ (the latter
two conditions imply each other).  If the $N=2$ theory is unitary one can
always choose representatives of the cohomology that are twisted chiral primary
states~\cite{LVW,DVV}.  In our case the theory is clearly nonunitary, since
the twisted ghost sector has $\hat c =-3$, and this will not always be
possible.

The second complication is that, as we have stressed, the TCFT of the NSR
string {\it must\/} be defined with the $(\beta,\gamma)$ ghosts bosonized, so
one should bear in mind some subtleties of the bosonization
procedure~\cite{FMS}.  Recall that the constant zero mode of $\xi$ is not in
the
Hilbert space of the original
($\beta,\gamma$) system.  This allows one to define the
nontrivial BRST-invariant ``picture-changing'' operator
$X=\left\{ Q \, , \, \xi \right\}$, and its inverse $Y$.
In the full Hilbert space of the bosonized theory all states that are closed
under $Q$ are exact~\cite{nocoho}, since the identity operator
$1 = X Y = \left\{ Q \, , \, \xi Y \right\}$ is exact.
The theory can therefore be nontrivial only on the reduced Hilbert space
without $\xi_0$.  The BRST cohomology in this reduced space consists of copies
of all the states of the NSR string repeated exactly once at each picture
number~\cite{coho}, as one would expect.  In the NSR string one regards states
in different pictures
as being equivalent, and one can transform between them using $X$ and $Y$.  In
the TCFT approach this equivalence is somewhat trickier, since $X$ and $Y$ have
nontrivial OPE's with the generators of the topological algebra.

Because of all this, it is instructive to examine the explicit forms of NSR
states in various pictures~\cite{FMS}.  In the NSR string, all amplitudes can
be---and generally are---calculated using only the vertex operators
$V_{(0)}$ and $V_{(-1)}$ in the NS sector and $V_{(1/2)}$ and $V_{(-1/2)}$ in
the R sector.  (Here the picture number of the operator is given in the
parenthesis.  Note that while the picture current
$J_P(z)$ is not BRST invariant, its charge $P$ is a good operator which
commutes with the entire topological algebra.  One can therefore use it, or
equivalently total ghost number, to distinguish distinguish between states in
different pictures.)  All these operators are annihilated by the $b_n$'s, and
so satisfy the first condition of states in a TCFT.  In fact, the
relative cohomology operators $V_{(0)}$, $V_{(-1)}$ and $V_{(-1/2)}$ satisfy
all the conditions for twisted chiral primary states.
(States in the absolute cohomology can not be twisted chiral primary states,
which must be annihilated by $b_0$.)  The operators $V_{(1/2)}$,
however, do not quite give twisted chiral primary states, since they contain a
piece that is not annihilated by $J_1$ and $Q_1$.  The only redeeming feature
of $V_{(1/2)}$ is that this piece does not contribute to amplitudes.
However, if one ventures into
other pictures, one finds that the situation deteriorates rapidly.  In
particular, one can show that there is {\it no\/} $P=-2$ operator $V_{(-2)}$
that is annihilated by the $b_n$'s\footnote{Such an operator can
contain at most one factor of $c$.  Then to have the correct R--parity all
extra $b$'s must be accompanied by $\eta$'s, and having $P=2$ fixes the power
of ${\rm e}^\phi$.  By dimensional analysis the only such operator possible is
$V_{(-2)} \mathrel{\mathop{\stackrel?=}} c\,{\rm e}^{-2 \phi} \, \tilde V_m$
($V_m$
and $\tilde V_m$ being the components of the superspace matter vertex), but
this is $ \{ \, Q \,,\, -2 c\,\partial^2 \xi {\rm e}^{-3 \phi} \, V_m \,\}$.}.
What is even worse is that the same argument shows that
any $V_{(-2)}$ must contain some $\xi_{-n}$, and
$\left[ J_n \, , \,  V_{(-2)}  \right]$
is then not in the Hilbert space of the theory, since it contains $\xi_0$.
We thus see that unless one sticks to the standard pictures, which is not at
all justified in the TCFT approach,
not only is it impossible to restrict oneself to
chiral primary states, but sometimes the very action of the $N=2$ algebra is
ill-defined.

Nevertheless, the theory does retain some features of
topological theories, since $T(z)$ and $Q(z)$ are exact~(\ref{exact}).  It is
interesting to try to see exactly how much of the structure of topological
theories is
retained in the NSR string, even if one has more the topological nature of the
bosonic string~\cite{DVV} than of the twisted minimal $N=2$ theories.  Since
all the operators in the cohomology must be dimensionless, one does have (an
infinite-dimensional) ring structure.  In usual NSR backgrounds, all relative
cohomology states in all pictures, which we shall denote $V_{(p)\,i\, \ldots}$,
have R--parity 1, both in the NS and in the R sectors.   The (remaining) states
$V_{(q)\,a\,\ldots}$ in the absolute cohomology have R--parity 2\footnote{Here
we do not worry about exotica such as non-critical NSR strings.  However, there
are exceptional discrete states in the cohomology at zero momentum, in
particular $c \, \eta$ which will play some role later.}.   Therefore,
the only nonvanishing products in the ring are:
\begin{equation}
\label{ring}
V_{(p)\,i} \cdot V_{(q)\,j} \, \sim \, {c_{ij}}^a(0) \; V_{(p+q)\,a} \,
+ \, \hbox{exact}\ .
\end{equation}
One can also
define a metric from the $2$-point function:
\begin{equation}
\label{2pt}
\eta_{ia}(0) = \left \langle \xi(z) \, V_{i}(z_1) \, V_{a}(z_2) \,
 \right\rangle \ ,
\end{equation}
in which the $\xi(z)$ is needed to soak up the zero mode $\xi_0$, or
equivalently to restrict the path integral to being only over the reduced
Hilbert space of the theory.  The metric does not depend on the positions of
the operators.  It is simply the usual metric in the NSR string, and acts
between relative cohomology states and absolute cohomology states.  We have
written ${c_{ij}}^a(0)$ and $\eta_{ia}(0)$, since in a usual topological theory
one would like to be able to generalize the theory by deforming the lagrangian
with two-form operators.  Unfortunately, in a real string one does not
know how to deform the theory in this way.  One
can think of the sigma-model approach to the string as a partial implementation
of such deformations, but it is not really adequate for the topological
approach.  In fact even the Taylor expansions of $\eta_{ia}$ and ${c_{ij}}^a$
can not be defined.  For example, $\partial_{i_1 \ldots i_l} \,\eta_{ia}(0)$
would come from amplitudes such as~(\ref{2pt}) with extra integrated
vertex operator insertions, but such amplitudes diverge, since one has not
fixed the M\"obius invariance of the sphere.  (And basically, there are never
sensible amplitudes involving absolute cohomology states in string theory.)

What are well defined in the topological approach are the amplitudes of the
relative cohomology states, as was seen for the
bosonic string in
ref.~\cite{DVV}.  The first issue in calculating amplitudes is to see
which ones are allowed by the (anomalously) conserved charges, and by the
cancellation of the zero modes of the fields in the theory.  In our case we
have R--parity, with background charge $-2$, and picture number, with
background
charge $1$.  Since one always needs an insertion of $\xi$, which has R--parity
$-1$ and $P=1$, one sees that the remaining operators in the
amplitude on the sphere must have total R--parity $3$  and $P=-2$.
The only way to fix the
R--parity is to have three zero-form relative cohomology operators, thus also
fixing the M\"obius invariance.  This give us the correlation functions
\begin{equation}
\label{3pt}
\left \langle \xi(z) \, V_{i}(z_1) \, V_{j}(z_2) \, V_{k}(z_3) \,
 \prod_{l=1}^{n} \, \int d^2 z_l \, V_{i_l} (z_l) \, \right\rangle \ ,
\end{equation}
which are the standard amplitudes of the NSR string.  As is usual in string
theory, the correlation functions are topological, in that they depend neither
on $z$ nor on the $z_i$'s,
and~(\ref{3pt}) can be thought of as the TCFT
amplitude $\partial_{i_1 \ldots \, i_n} \, c_{ijk}(0)$.  The total
picture charges of the $V$'s should add up to $-2$; clearly this can
be achieved using only the standard $-1 \le P \le  1/2$ pictures.
Note also that there
are no amplitudes with four or more zero-form operators in the theory, since
the R--parity can not then be canceled.   This means that
the factorization relations of the type found in topological matter
theories~\cite{DVV} are trivial here.

At higher genus one must couple the theory to topological gravity, in order to
have sensible amplitudes integrated over the moduli space of the Riemann
surfaces.  Then, as is usual in topological theories, one must insert $3 g-3$
integrals of  Beltrami differentials multiplied by $B$'s into the
amplitudes~\cite{DVV}.  In our case the $b$'s have R--parity $3-3g$ and
no picture charge, so the
remaining operators in the amplitude must have total R--parity $g$, and
$P=g-2$.  The usual expression for NSR string amplitudes involves only
$2g-2$ extra insertions of picture-changing operators $X$,
together with some integrated two-form vertex operators~\cite{v2-pict}.
None of these carries R--parity, which therefore can not be balanced.
However the $X$ insertions are basically heuristic translations into
the bosonized language of the fact that one needs $2g-2$ insertions of
$\delta(\beta) \cdot G_m$'s into the NSR path integral, to soak up the zero
modes of $\beta$ and the super-Teichm\"uller parameters.
If one calculates NSR amplitudes
directly in the bosonized language, one sees that in addition one has to insert
$g$ $\eta$'s, integrated around each $a$--cycle of the Riemann
surface~\cite{Carow}.  These insertions ensure that if the Riemann surface is
cut,
only states
in the restricted Hilbert space propagate to the rest of the
surface~\cite{Carow}.  They are BRST invariant, properly take care of the $g$
zero-modes of the $\eta$ field, and in addition provide us with the extra
R--parity we need (and also picture charge $-g$).  Also, in order to avoid
an infinite overcounting of states,
one should restrict the picture number $P_\gamma$ flowing through any
$a$--cycle, $P_\gamma = \oint_{{\cal A}_\gamma} \! J_P$, to be some (arbitrary)
value $p_\gamma$~\cite{Carow}.  As we have stated before, $P$ is also BRST
invariant.  All of these considerations fit nicely into the topological
framework, giving us the final formula for higher genus amplitudes:
\begin{equation}
\label{g>1}\int_{\cal M} \, \left \langle \xi(z) \cdot
 \prod_\alpha^{3g-3} \, \int \mu^\alpha b \cdot
 \prod_{\beta=1}^{2g-2} \, X(z_\beta) \cdot
 \prod_{\gamma=1}^{g} \, \delta_{P_\gamma \, , \, p_\gamma} \,
       \oint_{{\cal A}_\gamma} \eta \cdot
  \prod_{l=1}^{n} \, \int d^2 z_l \, V_{i_l} (z_l) \, \right\rangle \ .
\end{equation}
The vertex operator
insertions need to have total picture number $P=0$.  The amplitudes do not
depend on the positions of the insertions of $\xi$ and of the $X$'s, or on the
picture charges $p_\gamma$, and agree with those of the NSR string.

In conclusion, we have seen that the NSR string in any background has a closed
topological $N=2$ algebra, given by the generators of~(\ref{ops}).
The ($\beta,\gamma$) ghost system {\it must\/}
be bosonized in order to define the R--parity
current of the algebra, which, in contrast
to previous constructions of this type~\cite{alltop1,alltop2},
does not depend on the matter sector of the theory.
In the pictures that are usually considered, the states of the
string are (in the case of $V_{(1/2)}$, almost) chiral primary states of the
$N=2$.  However, this is not the case in arbitrary pictures, and the action of
the $N=2$ generators in some pictures is not even well-defined.  This may not
be too important, in that even in usual topological conformal field theories
the twisted $N=2$ algebra itself does not play much of a role.  The simplicity
of these theories comes mainly from the fact that there are only a finite
number of states, leading to a finite ring structure, but this can never happen
in any physical string where an entire infinite spectrum must be represented in
the cohomology.

As was seen in the case of the bosonic string~\cite{DVV}, the NSR string
does have many of the features of a topological
conformal field theory.  In particular, the topological approach does
give the correct states and amplitudes of the string.
Clearly, the
most interesting questions remaining are whether writing
apparently nontopological string theories in a topological way can be used
for something practical, despite the fact that these theories have an
infinite number of states, and whether one can find a deeper
topological meaning of these theories.

\vskip 1 cm

\parskip 0 pt plus 1 pt
\parindent 0 em

I would like to thank Shimon Yankielowicz,  Yaron Oz and Cobi Sonnenschein for
many helpful discussions, and Yaron Oz for leading me to consider this problem.

\end{document}